\documentclass[fleqn,usenatbib]{mnras}

\usepackage{newtxtext,newtxmath}

\usepackage[T1]{fontenc}

\DeclareRobustCommand{\VAN}[3]{#2}
\let\VANthebibliography\thebibliography
\def\thebibliography{\DeclareRobustCommand{\VAN}[3]{##3}\VANthebibliography}

\usepackage{graphicx}	
\usepackage{amsmath, color}	

\title[$S_8$ increases with redshift in $\Lambda$CDM]{$S_8$ increases with effective redshift in $\Lambda$CDM cosmology}

\author[S. A. Adil et al.]{
S. A. Adil,$^{1}$
\"{O}. Akarsu,$^{2}$
M. Malekjani,$^{3}$ 
E. \'O Colg\'ain \thanks{E-mail: eoin.ocolgain@atu.ie},$^{4}$ 
S. Pourojaghi,$^{5}$ 
A. A. Sen $^{6}$ 
\newauthor
and M. M. Sheikh-Jabbari $^{5}$
\\
$^{1}$Department of Physics, Jamia Millia Islamia, New Delhi - 110025, India\\
$^{2}$Department of Physics, Istanbul Technical University, Maslak 34469 Istanbul, Turkey\\
$^{3}$Department of Physics, Bu-Ali Sina University, Hamedan 65178, Iran\\
$^{4}$Atlantic Technological University, Ash Lane, Sligo, Ireland\\
$^{5}$School of Physics, Institute for Research in Fundamental Sciences (IPM), P.O.Box 19395-5531, Tehran, Iran\\
$^{6}$Centre for Theoretical Physics, Jamia Millia Islamia, New Delhi - 110025, India
}

\date{Accepted XXX. Received YYY; in original form ZZZ}

\pubyear{2023}

\begin{document}
\label{firstpage}
\pagerange{\pageref{firstpage}--\pageref{lastpage}}
\maketitle

\begin{abstract}
Hubble constant $H_0$ and weighted amplitude of matter fluctuations $S_8$ determinations are biased to higher and lower values, respectively, in the late Universe with respect to early Universe values inferred by the Planck collaboration within flat $\Lambda$CDM cosmology. If these anomalies are physical, i.e. not due to systematics, they naively suggest that $H_0$ decreases and $S_8$ increases with effective redshift. Here, {subjecting matter density today $\Omega_{m}$ to a prior, corresponding to a combination of Planck CMB and BAO data, we perform a consistency test of the Planck-$\Lambda$CDM cosmology and show that $S_8$ determinations from $f \sigma_8(z)$ constraints increase with effective redshift. Due to the redshift evolution, a $\sim 3 \sigma$ tension in the $S_8$ parameter with Planck at lower redshifts remarkably becomes consistent with Planck within $1 \sigma$ at high redshifts.} This provides corroborating support for an $S_8$ discrepancy that is physical in origin. {We further confirm that the flat $\Lambda$CDM model is preferred over a theoretically ad hoc model with a jump in $S_8$ at a given redshift. In the absence of the CMB+BAO $\Omega_m$ prior, we find that $> 3 \sigma$ tensions with Planck in low redshift data are ameliorated by shifts in the parameters in high redshift data.} Results here and elsewhere suggest that the $\Lambda$CDM cosmological parameters are redshift dependent. Fitting parameters that evolve with redshift is a recognisable hallmark of model breakdown. 
\end{abstract}




\section{Introduction} 
Modern cosmology is a pursuit where one theoretically guesses a function (or model) and compares it to another \textit{a priori} unknown function, the Hubble parameter $H(z)$, extracted observationally from Nature. Given the possibilities, the guess is invariably wrong, but existing cosmological data can mask disagreement until precision improves. Viewed historically, concordance  Lambda cold dark matter ($\Lambda$CDM) cosmology emerged when a spatially flat, homogeneous \& isotropic Universe, filled exclusively with radiation and (pressure-less) matter, required a dark energy sector to explain observations~\cite{SupernovaSearchTeam:1998fmf, Perlmutter_1999}. Further improvements in data in recent years have unveiled anomalies in the Hubble constant $H_0 := H(z=0)$~\cite{Planck:2018vyg, Riess:2021jrx, Freedman:2021ahq, Pesce:2020xfe, Blakeslee:2021rqi, Kourkchi:2020iyz}, the weighted amplitude of matter fluctuations $S_8 := \sigma_8 \sqrt{\Omega_{m}/0.3}$~\cite{Planck:2018vyg, Heymans:2013fya, Joudaki:2016mvz, DES:2017qwj, HSC:2018mrq, KiDS:2020suj, DES:2021wwk}, the lensing parameter $A_{\textrm{lens}}$ and/or curvature $\Omega_k$~\cite{Planck:2018vyg, Addison:2015wyg, Handley:2019tkm, DiValentino:2019qzk}, the late-time integrated Sachs-Wolfe (ISW) effect~\cite{Granett:2008ju, DES:2016zxh, DES:2018nlb, Kovacs:2021mnf}, and high redshift galaxies that seemingly defy $\Lambda$CDM expectations~\cite{Adams:2022, Labbe:2022, Castellano:2022, Naidu:2022}. See~\cite{DiValentino:2021izs, Perivolaropoulos:2021jda, Abdalla:2022yfr} for reviews of $\Lambda$CDM anomalies. It is plausible that the (flat) $\Lambda$CDM cosmological model is breaking down. What remains is to confirm this diagnosis. 

To that end, a simple insight comes directly to us from the Friedmann equations; $H_0$ is by construction an integration constant. In other words, it is \textit{theoretically} a constant within the Friedmann-Lema\^itre-Robertson-Walker (FLRW) framework on which $\Lambda$CDM is established by assuming spatial flatness. Nevertheless, it is \textit{observationally} a constant only when one has the correct cosmological model~\cite{Krishnan:2020vaf, Krishnan:2022fzz}. This statement is true not only for $H_0$, but also for other model parameters that are integration constants, such as matter density parameter (today) $\Omega_m:=\rho_{m0}/3H_0^2$ ($\rho_{m0}$ being the matter energy density today) in the $\Lambda$CDM model. Thus,  if the $\Lambda$CDM model is breaking down, as all cosmological models must at some precision for the reason outlined above, one expects $H_0$, $\Omega_m$, etc, to evolve with effective redshift. There are now numerous observations suggesting that $H_0$ evolves,  more accurately decreases with effective redshift~\cite{Wong:2019kwg, Millon:2019slk, Krishnan:2020obg, Dainotti:2021pqg, Dainotti:2022bzg, Colgain:2022nlb, Colgain:2022rxy, Malekjani:2023dky} (see also~\cite{Hu:2022kes, Jia:2022ycc}). Likewise there are claims of $\Omega_m$ increasing with effective redshift~\cite{Colgain:2022nlb, Colgain:2022rxy, Malekjani:2023dky} (see also 
\cite{Risaliti:2018reu, Lusso:2020pdb, Yang:2019vgk, Khadka:2020vlh,Khadka:2020tlm, Khadka:2021xcc, Pourojaghi:2022zrh, Pasten:2023rpc}). Furthermore, it has been noted in~\cite{Colgain:2022nlb, Colgain:2022rxy, Malekjani:2023dky} that evolution of $H_0$ and $\Omega_m$ within $\Lambda$CDM are anti-correlated with each other. Note that if $\Omega_m$ evolves with effective redshift, it is unlikely that $S_8$ is a constant, because $\sigma_8$ conspiring to balance evolution in $\Omega_{m}$ represents a contrived or unnatural scenario. 

Here we focus on the $S_8$ discrepancy~\cite{Planck:2018vyg, Heymans:2013fya, Joudaki:2016mvz, DES:2017qwj, HSC:2018mrq, KiDS:2020suj, DES:2021wwk}. Taken at face value, this anomaly says that $S_8$ as inferred by the Planck collaboration (high redshift inference) has a larger value than galaxy surveys (low redshift measurement).
This can be independently verified with growth rate data, in particular measurements of $f \sigma_8 (z)$, which are independent of galaxy bias. Concretely, we will assume expressions that are valid for $\Lambda$CDM cosmology and fit $f \sigma_8 (z)$ constraints in a single bin of increasing effective redshift. However, since growth rate data suffers from large fractional errors, we will leverage a working assumption in modern cosmology that $\Omega_m$ is {tightly constrained} to elucidate the trend. This means that the increases observed in $S_8$ are driven by increases in $\sigma_8$. Interestingly, {this effect is also evident when one compares $\sigma_8$ constraints from number counts of galaxy clusters identified through the Sunyaev-Zeldovich effect at low redshifts with    $\sigma_8$ constraints from Lyman-$\alpha$ spectra at high redshifts~\cite{Esposito:2022plo}. Moreover, CMB lensing, an observable most sensitive to redshift ranges $z \in [0.5, 5]$ and peaking in sensitivity at $z \sim 2$, recovers the Planck result \cite{ACT:2023dou, ACT:2023kun}. This seemingly constrains any ``$S_8$ tension'' to the late Universe \cite{ACT:2023ipp} \footnote{{Intriguingly, higher redshift observables may prefer lower values of $\sigma_8$ relative to Planck \cite{Miyatake:2021qjr, Alonso:2023guh}. $\Omega_m$ is poorly constrained at high redshifts.}}.} Admittedly, $\Omega_m$ may not be a constant~\cite{Colgain:2022nlb, Colgain:2022rxy}, but this is currently a fringe viewpoint. While $S_8$ or $\sigma_8$ tension in growth rate data is well studied in the literature \cite{Macaulay:2013swa,Battye:2014qga, Nesseris:2017vor, Kazantzidis:2018rnb, Skara:2019usd, Quelle:2019vam, Li:2019nux, Benisty:2020kdt, Nunes:2021ipq}, our goal in this letter is to explore redshift evolution of $S_8$ within $\Lambda$CDM in the late Universe under standard assumptions. 

\section{Warm Up} 
\begin{figure}
\includegraphics[width=80mm]{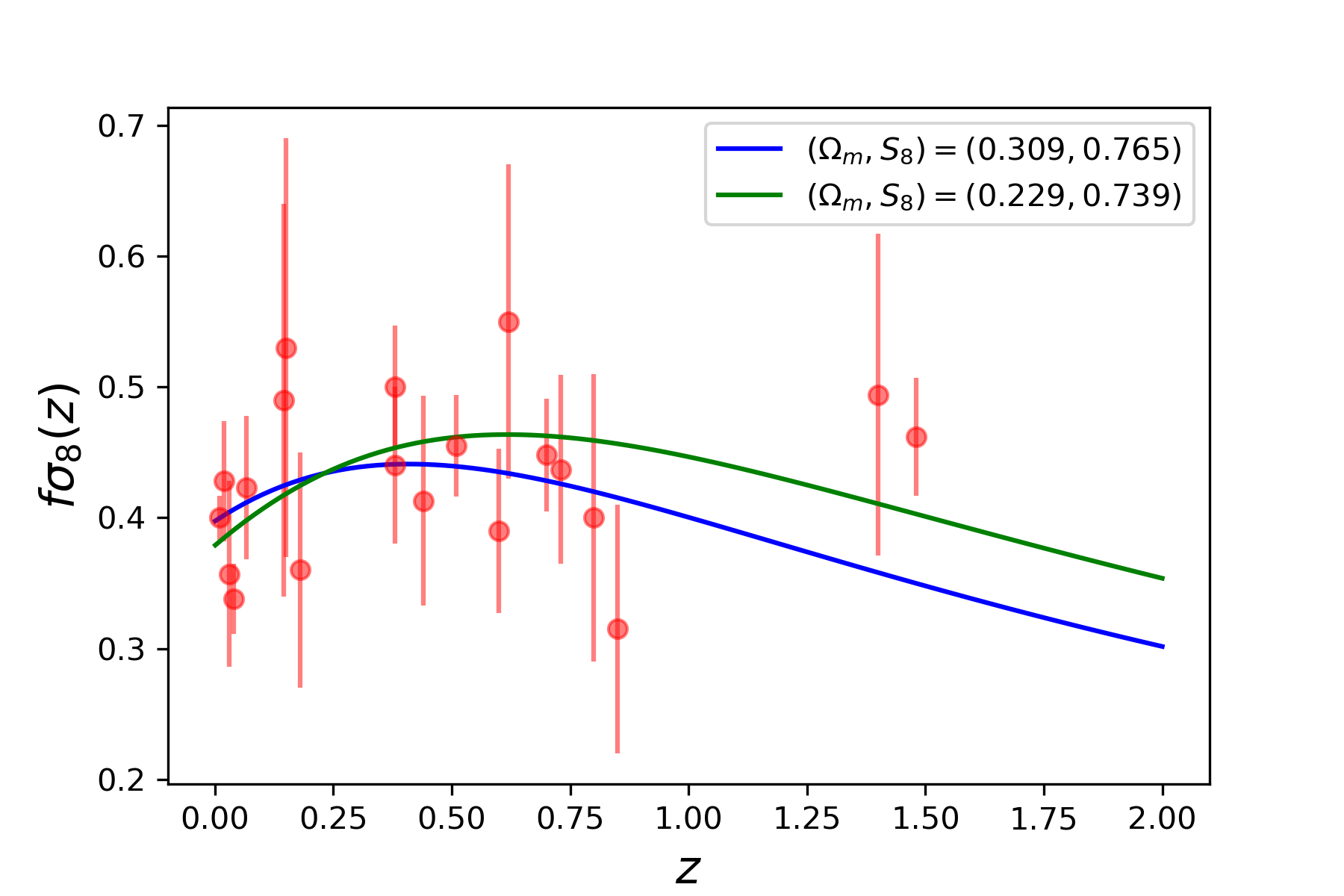} 
\caption{$f \sigma_8 (z)$ constraints in red
with best fit values of $(\Omega_m, S_8)$ both with (blue) and without (green) a Gaussian prior $\Omega_m = 0.3111 \pm 0.0056$.}
\label{fig:dragan_data}
\end{figure}

To get oriented, we impose a Gaussian prior, $\Omega_{m} = 0.3111 \pm 0.0056$, which arises from combining Cosmic Microwave Background (CMB), galaxy, quasar and Lyman-$\alpha$ baryon acoustic oscillation  (BAO) constraints~\cite{Planck:2018vyg, BOSS:2016wmc, Hou:2020rse, Neveux:2020voa}. {In line with standard practice, we assume there is no discrepancy between CMB and BAO on $\Omega_m$ inferences, i. e. that the key success of the $\Lambda$CDM model is not undermined.} The CMB+BAO prior is needed to compensate for the relatively low quality of the  growth rate data, which will be further reduced {by the removal of low redshift data}. Compared to Planck \cite{Planck:2018vyg}, our constraint is marginally more constraining, but remains representative of a Planck prior. In the absence of Gaussian priors, we employ {uninformative uniform priors $\Omega_m \sim \mathcal{U} (0, 1)$ and $\sigma_8 \sim \mathcal{U}(0, 1.5)$ throughout}. 
We combine this prior with 20 measurements of $f \sigma_8 (z)$ from peculiar velocity and redshift-space distortion (RSD) data~\cite{Said:2020epb, Beutler:2012px, Huterer:2016uyq, Boruah:2019icj, Turner:2022mla, Blake:2011rj, Blake:2013nif, Howlett:2014opa, Okumura:2015lvp, Pezzotta:2016gbo, eBOSS:2020yzd}, as compiled recently in~\cite{Nguyen:2023fip}. {Modulo the removal of low redshift data, we follow the methodology of ~\cite{Nguyen:2023fip}.} We illustrate the constraints in Fig.~\ref{fig:dragan_data} 
in red. We will be interested in the combination 
\begin{equation}
\label{s8}
S_8 := \sigma_{8} \sqrt{\Omega_m/0.3}, 
\end{equation}
where the constant $\sigma_8$ is the amplitude of matter fluctuations in spheres of $8 h^{-1}$ Mpc with $h:=H_0/100\, {\rm km\,s}^{-1}{\rm Mpc}^{-1}$. Following~\cite{Wang:1998gt}, we introduce the matter density parameter, 
\begin{equation}
\Omega(z) := {\frac{\Omega_m(z)}{H(z)^2/H_0^2}}  = \frac{\Omega_{m} (1+z)^3}{1- \Omega_{m} + \Omega_{m} (1+z)^3}, 
\end{equation}
thereby allowing us to obtain 
\begin{equation}
\label{fs8}
f \sigma_8 (z)  = \sigma_{8} \, \Omega^{\frac{6}{11}}(z) \exp \left( - \int_0^{z} \frac{\Omega^{\frac{6}{11}}(z^{\prime})}{1+z^{\prime}} \textrm{d} z^{\prime} \right).  \end{equation}
{As explained in~\cite{Wang:1998gt}, these expressions are a  valid approximation for $\Lambda$CDM. Concretely, we checked that the fractional error between the approximation and the exact expression for $f \sigma_8(z)$ based on the hypergeometric function ${}_2 F_1$ (see \cite{Nesseris:2017vor}) is greatest at $1 \%$ at $z=0$. In the range $0.35 \lesssim z \lesssim 2$, the fractional error is less than $0.2 \%$. This uncertainty is negligible compared to the observational uncertainties in Fig.~\ref{fig:dragan_data}. Working with an approximation may seem obsolete, but it allows us to shed light on evolution that is hidden in the analysis of~\cite{Nguyen:2023fip} \footnote{{In \cite{Nguyen:2023fip} it is assumed that there is no evolution in cosmological parameters across the $f \sigma_8(z)$ constraints.}}, where the same approximation and data are employed.} For all redshifts $z$, the function $\Omega(z)$ is bounded $ \Omega_{m} \leq \Omega (z)  <1$. In particular, $\Omega \rightarrow 1$ as $z \rightarrow \infty$, so that $\Omega_m$ dependence drops out as an overall factor in (\ref{fs8}) at high redshifts, but some knowledge of $\Omega_m$ is retained through the integral. {Thus, one expects $\Omega_m$ to be poorly constrained at higher redshifts. As we show in the appendix, one risks encountering a degeneracy between the fitting parameters $(\Omega_m, \sigma_8)$ that is difficult to cleanly break with exclusively high redshift binned data. Mathematically, one is guaranteed to run into a problem constraining $\Omega_m$ in high redshift bins, so it is prudent to impose an $\Omega_m$ prior. Later we will relax the prior and comment on the changes.}

\begin{figure}
\includegraphics[width=80mm]{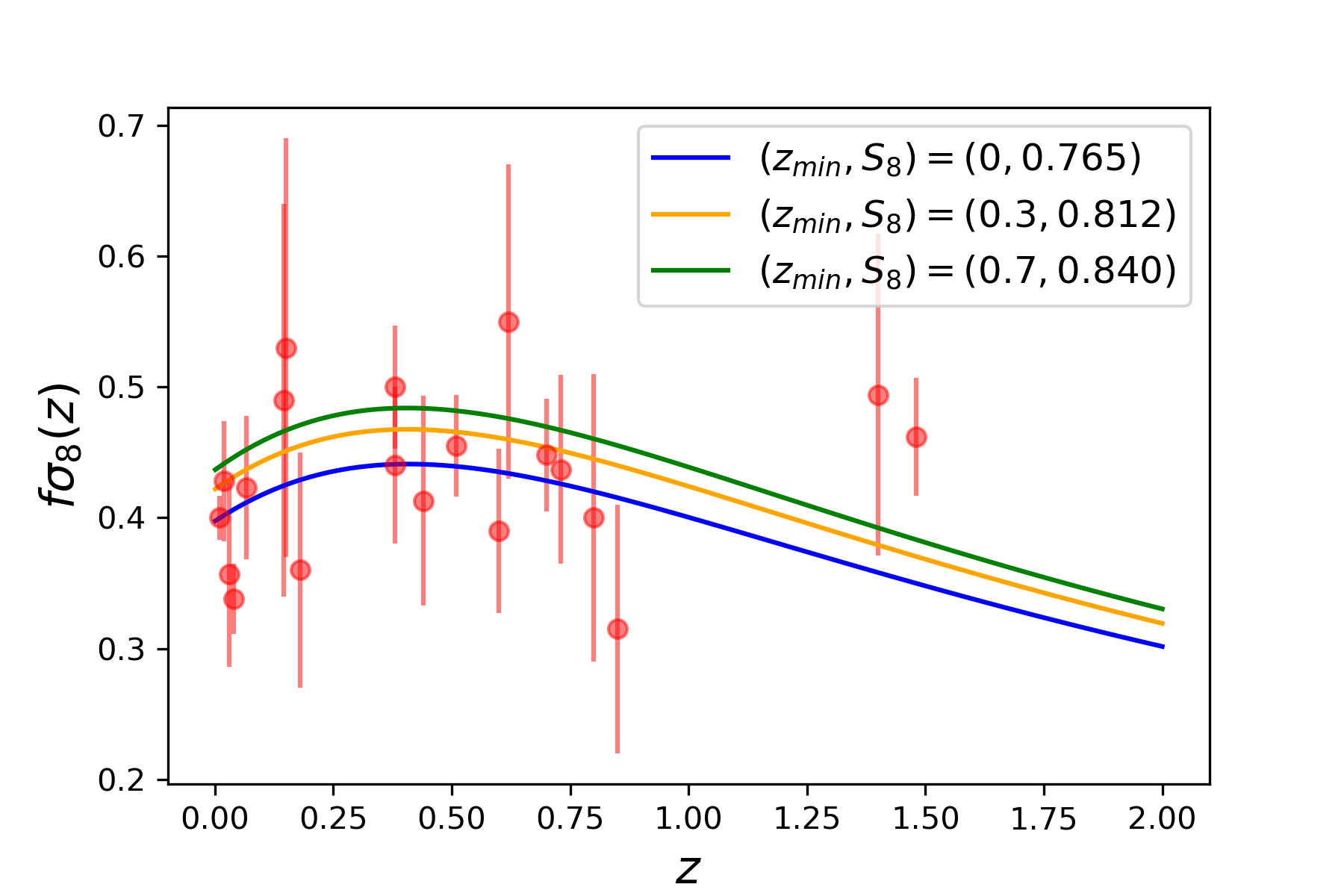} 
\caption{Same as Fig.~\ref{fig:dragan_data} with Gaussian prior $\Omega_m = 0.3111 \pm 0.0056$ and data with  $z \leq z_{\textrm{min}}$ removed. $S_8$ increases with effective redshift.}
\label{fig:zmin}
\end{figure}

The green curve in Fig.~\ref{fig:dragan_data} denotes the best fit of the $\Lambda$CDM model to the 20 data points, whereas the blue curve documents the effect of introducing the Gaussian prior on $\Omega_{m}$. It is worth noting that this increases the $S_8$ inference and visibly worsens the fit to the $f \sigma_8 (z)$ constraints at $ z \approx 0$ and $z \approx 1.5$, thereby underscoring the tension between the $\Omega_{m}$ prior and the lower value of $\Omega_m$ preferred by $f \sigma_8 (z)$ data. The blue curve worsens the fit to $f \sigma_8 (z)$ data by $\Delta \chi^2 \approx 4.9$. The $f \sigma_8 (z)$ constraints are weak while the Gaussian prior is strong. Therefore, the prior effectively fixes $\Omega_{m}$, so that the only parameter being fitted is $\sigma_8$. As is clear from (\ref{fs8}), one is simply fitting the scale of the $f \sigma_8(z)$ function, whereas the functional form is fixed. In Fig.~\ref{fig:zmin} we see the effect of removing $f \sigma_8 (z)$ constraints below $z = z_{\textrm{min}}$. We see that with $\Omega_{m}$ constrained through the prior, $\sigma_8$ increases leading to larger values of $S_8$. As $S_8$ increases, the curves visibly provide a better fit to high redshift $f \sigma_8 (z)$ constraints. This increase in $S_8$ is driven by the two high redshift data points and is expected, {since it is visible in the raw data, i. e. no data analysis is required. Note, the eBOSS data point at $z \sim 1.5$ is re-analysed in \cite{Brieden:2022lsd} with different methodology and the central point evolves little. There is nothing to suggest this data point is not robust}.   

\begin{figure}
\includegraphics[width=80mm]{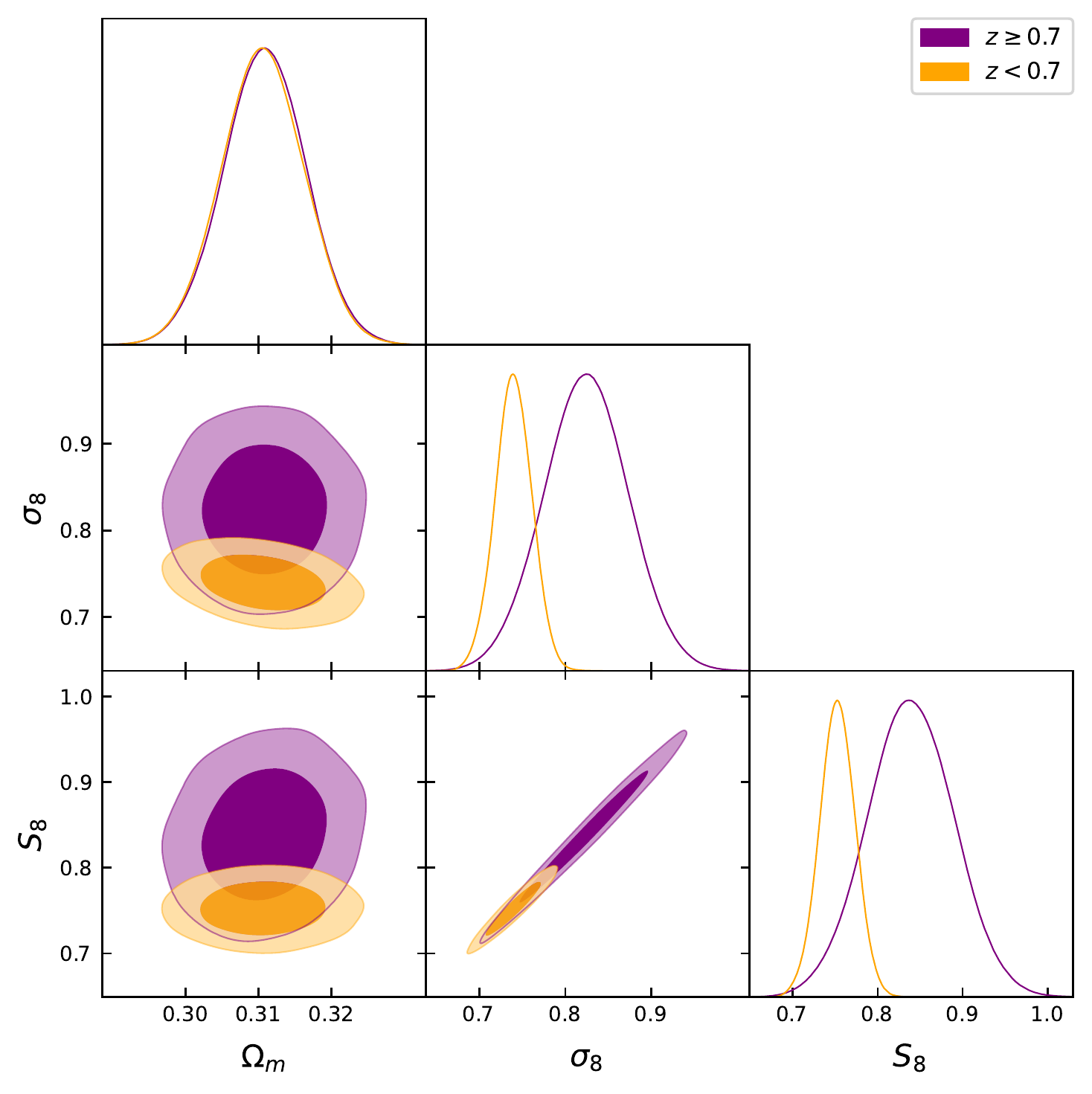} 
\caption{Marginalisation of parameters $(\Omega_m, \sigma_8, S_8)$ with Gaussian prior $\Omega_m = 0.3111 \pm 0.0056$ and compilation of $f \sigma_8(z)$ data
split at $z = 0.7$. $S_8$ is reconstructed from $(\Omega_m, \sigma_8)$ MCMC chains. The discrepancy in the $S_8$ plane between low and high redshift subsamples is $1.6 \sigma$.}
\label{fig:dragan_om_prior}
\end{figure}

We next split the $f \sigma_8(z)$ data at $z= 0.7$ into a low redshift sample of 14 data points and a high redshift sample of 6 data points. This choice is arbitrary, but it is guided by Fig.~\ref{fig:dragan_data} and the intuition that evolution in $S_8$ is expected once we see evolution in $H_0$~\cite{Wong:2019kwg, Millon:2019slk, Krishnan:2020obg, Dainotti:2021pqg, Dainotti:2022bzg, Colgain:2022nlb, Colgain:2022rxy, Malekjani:2023dky} and $\Omega_m$~\cite{Colgain:2022nlb, Colgain:2022rxy, Malekjani:2023dky} elsewhere. We impose the Gaussian prior on $\Omega_{m}$ to each sample and perform a Markov Chain Monte Carlo (MCMC) analysis. The result is shown in Fig.~\ref{fig:dragan_om_prior}, where in line with expectations from Fig.~\ref{fig:zmin}, the contours separate in the $\sigma_8$ or $S_8$ direction. {The contours are Gaussian, allowing us to directly compare $S_8 = 0.753^{+0.021}_{-0.020}$ $(z < 0.7)$ to $S_8 = 0.839^{+0.051}_{-0.050}$ $(z \geq 0.7)$ and conclude that the discrepancy in $S_8$ is $1.6 \sigma$. Furthermore, the former value is discrepant with the Planck value $S_8 = 0.832 \pm 0.013$~\cite{Planck:2018vyg} at $3.2 \sigma$. Throughout, we quote $1 \sigma$ confidence intervals. Note that the $1.6 \sigma$ shift occurs within the 20 data point sample, which simply assumes that BAO and CMB can consistently constrain $\Omega_m$. Observe also that the effect of the shift is to remove a $\sim 3 \sigma$ tension with Planck in the $S_8$ parameter.}

\begin{figure}
\includegraphics[width=80mm]{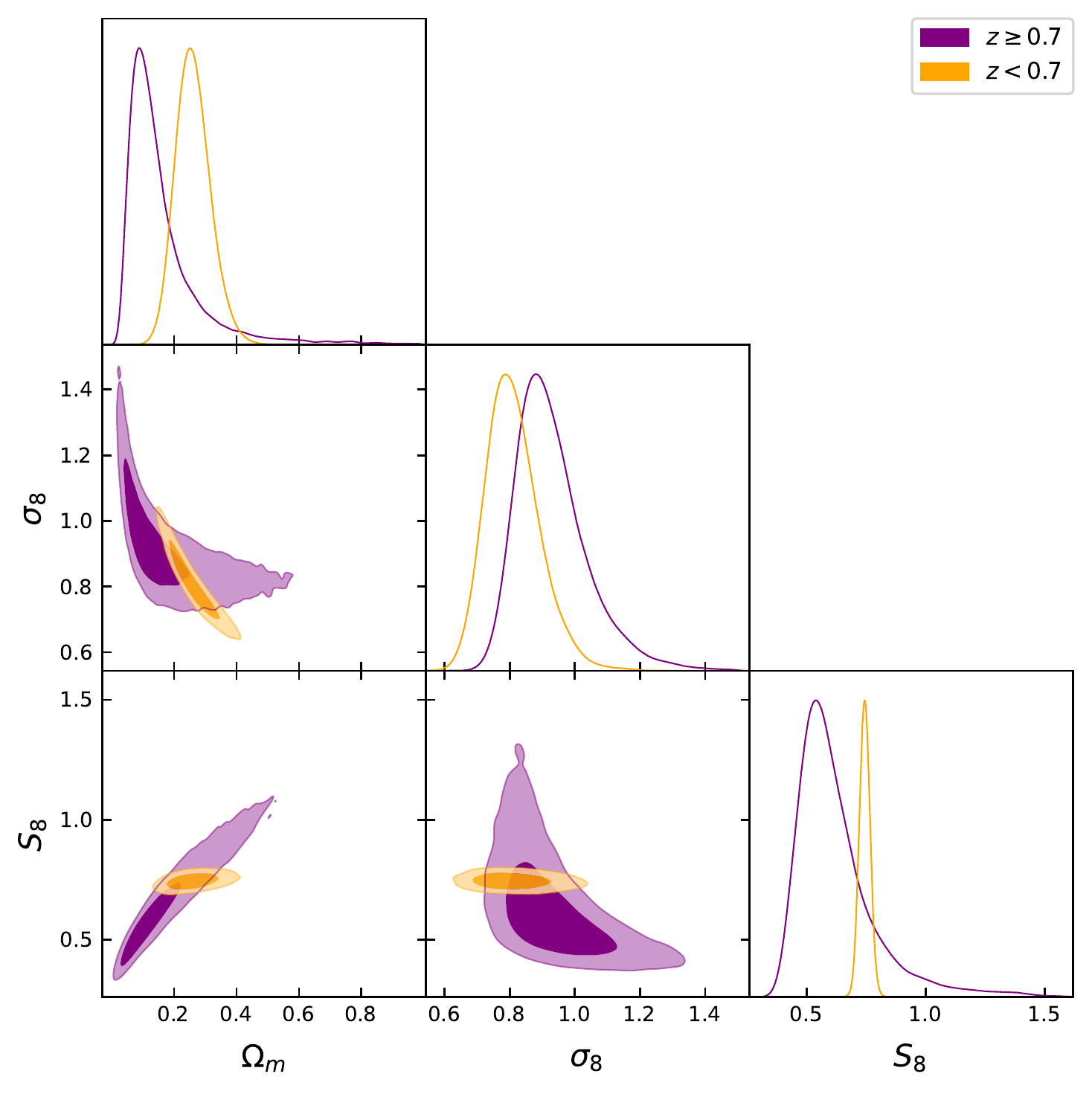} 
\caption{{Same as Fig. \ref{fig:dragan_om_prior} but without a $\Omega_m$ prior.}}
\label{fig:dragan_om_no_prior}
\end{figure}

\begin{figure}
\includegraphics[width=80mm]{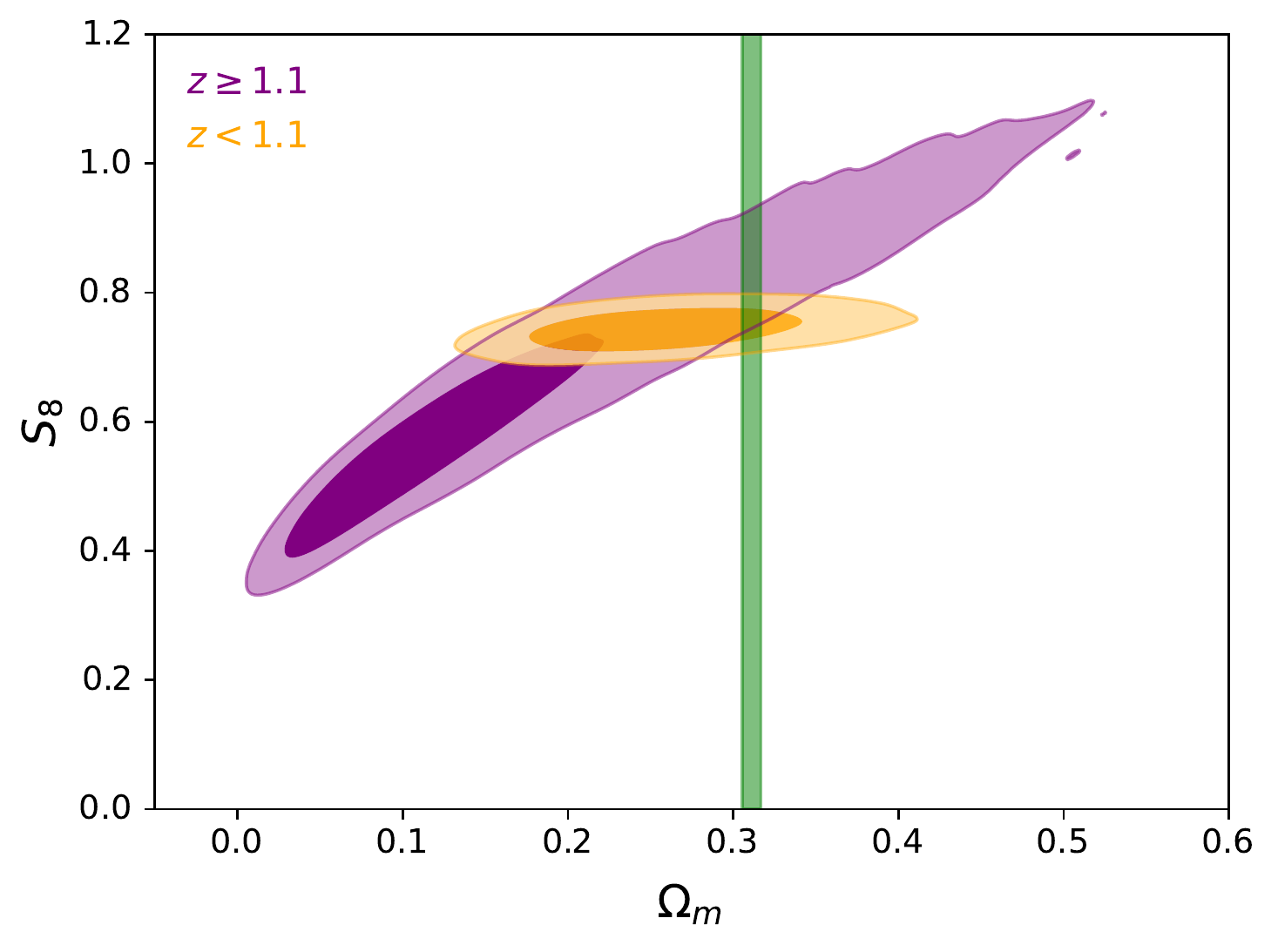} 
\caption{{The CMB+BAO prior $\Omega_m = 0.3111 \pm 0.0056$ intersects the confidence intervals at larger values of $\Omega_m$ resulting in a larger value of $S_8$ in the high redshift ($z \geq 1.1$) sample. The prior is consistent with low and high redshift constraints within $2 \sigma$, so constraints can be safely combined.}}
\label{fig:prior_effect}
\end{figure}

{In Fig.~\ref{fig:dragan_om_no_prior}, we investigate the effect of removing the $\Omega_m$ prior. The low redshift sample leads to $(\Omega_m, \sigma_8)$ values consistent within $1 \sigma$ with Planck, $(\Omega_m, \sigma_8) = (0.257^{+0.058}_{-0.051}, 0.803^{+0.088}_{-0.073})$, however $S_8 = 0.744^{+0.022}_{-0.022}$ is discrepant at $3.4 \sigma$. We emphasise that we sample $(\Omega_m, \sigma_8)$, but reconstruct $S_8$ as a derived parameter from the MCMC chains. The higher redshift contours are shifted away from their Planck values by more than $1 \sigma$ throughout, $(\Omega_m, \sigma_8, S_8) = (0.124^{+0.117}_{-0.055}, 0.913^{+0.120}_{-0.084}, 0.585^{+0.172}_{-0.101})$, but remain consistent within $2 \sigma$. It is clear that lower values of $S_8$ are driven by lower than expected values of $\Omega_m$ in both low and high redshift subsamples. Note that caution is required when interpreting these shifts in the absence of an informative prior, as the high redshift sample is small with only 6 data points.
In Fig. \ref{fig:prior_effect}, we demonstrate the consistency between the $\Omega_m$ prior and the confidence intervals from low and high redshift samples in the $(\Omega_m, S_8)$-plane. The intersection of the prior $1 \sigma$ confidence interval explains the preference for the higher $S_8$ value from the high redshift subsample. Since all constraints are within $2 \sigma$, no objection to combining the constraints is foreseen. While removing the $\Omega_m$ prior makes a considerable difference, it is nevertheless true that we see less tension with Planck in the high redshift sample following shifts in cosmological parameters.} 

{As an aside, with an $\Omega_m$ prior, we note that the $\Lambda$CDM model is still preferred over any theoretically poorly motivated model with a jump in $\sigma_8$ at $z = 0.7$. The reduction in the minimum of the $\chi^2$, $\Delta \chi_{\textrm{min}}^2 = -2.59$ is not enough to overcome the penalty of introducing two additional parameters, an additional $\sigma_8$ and a redshift for the jump $z_{\textrm{jump}}$, even in the Akaike Information Criterion (AIC). Our observation here is essentially a microcosm of the $\Lambda$CDM tensions debate; despite seeing symptoms of a problem, here evolution in $S_8$ when $f \sigma_8(z)$ constraints are combined with CMB+BAO, it is a separate matter to produce a model that outperforms $\Lambda$CDM.} 

{Given the relatively small size of the $f \sigma_8(z)$ data set considered in this work, in the appendix we work with a larger data set of historical data. The shortcomings of this data set are explicitly discussed, but we arrive at similar conclusions. }

\section{Conclusions}
Either from the Friedmann equations directly, or the continuity equation and the assumption of pressure-less matter, both $H_0$ and $\Omega_m$ arise \textit{mathematically} as integration constants. Thus, consistency demands that both $H_0$ and $\Omega_{m}$ are \textit{observationally} constants. An important corollary is that CMB, BAO, Type Ia supernovae, etc, must provide consistent $\Omega_m$ constraints in $\Lambda$CDM cosmology. Similarly, {the fitting parameter $\sigma_8$ must be a constant, as it is related to the normalisation of the matter power spectrum}. Here, working within {the assumption that CMB and BAO consistently constrain $\Omega_m$}, we have shown that the $S_8 := \sigma_8 \sqrt{\Omega_{m}/0.3}$ parameter increases with effective redshift in the late Universe, $z \lesssim 2$. This evolution is compelling because it corroborates the tendency of cosmic shear to give lower values of $S_8$ than Planck-$\Lambda$CDM~\cite{Planck:2018vyg, Heymans:2013fya, Joudaki:2016mvz, DES:2017qwj, HSC:2018mrq, KiDS:2020suj, DES:2021wwk}. Moreover, it is not an isolated observation; {we note that i) $\sigma_8$ constraints from low redshift clusters are biased lower than high redshift Lyman-$\alpha$ spectra \cite{Esposito:2022plo} and ii) weak lensing $S_8$ constraints, sensitive to lower redshifts, are biased lower than CMB lensing results \cite{ACT:2023dou, ACT:2023kun, ACT:2023ipp} that are sensitive to higher redshifts.} In \cite{Esposito:2022plo, ACT:2023dou, ACT:2023kun, ACT:2023ipp}, systematics are a greater concern as one is comparing different observables, but here we are working with common $f \sigma_8(z)$ constraints throughout, {so systematics should be under greater control. The data largely comprises RSD but peculiar velocity constraints may be present at $z \sim 0$, e. g. \cite{Boruah:2019icj, Said:2020epb}}. However, these differences aside, we agree on the increasing $\sigma_8$ trend, {a trend that is evident from the raw data.}

It is instructive to recall the assumptions being made: 
\begin{enumerate}
\item Equation (\ref{fs8}) is a valid approximation for $\Lambda$CDM behaviour.
\item $\Omega_m$ in $\Lambda$CDM is approximately $0.3$.
\item Data sets and/or priors from independent data sets can be combined.
\item The data are correct. 
\end{enumerate}
If this set of assumptions is correct, then the increasing $\sigma_8/S_8$ trend follows. Assumption i) is standard \cite{Wang:1998gt} {and the theoretical uncertainty introduced is less than $1 \%$, more accurately $0.2 \%$ in the redshift range $0.35 \lesssim z \lesssim 2$.} Assumption ii) is a widely recognised observation. Assumption iii) is standard practice. Assumption iv) is the weakest, but the removal of data points requires justification. In other words, removing data points with the sole motivation of recovering a null result is unscientific.

Concretely, we presented analysis in a sample of 20 and 66 $f \sigma_8(z)$ {(see appendix for the latter)} constraints respectively. In both samples we find the same increasing $S_8$ trend, which is essentially a $\sigma_8$ trend, because $\Omega_m$ is subject to a strong prior. Since there is considerable survey overlap in Fig.~\ref{fig:leandros_prior}, the tension between low and high redshift is admittedly \textit{overestimated} at {$2.8 \sigma$} (Fig.~\ref{fig:leandros_prior}). That being said, if the 20 data points in Fig.~\ref{fig:dragan_data} are independent (Ref.~\cite{Nguyen:2023fip} assumes they are), then {there is a $1.6 \sigma$ shift in $S_8$ that warrants further study, especially since it seems to ameliorate tension with Planck}. One could identify other redshift ranges where the discrepancy between low and high redshift inferences is less, but such an exercise is meaningless. To stress test any sample against evolution, one needs to focus on splits that exacerbate the feature. Moreover, Fig. \ref{fig:leandros_data_trend} clearly demonstrates that evolution is present throughout the sample. {Despite the presence of evolution, we find that the $\Lambda$CDM model is still preferred over any theoretically ad hoc model with a jump in $S_8$. Finally, when removing the $\Omega_m$ prior, despite noticeable differences in cosmological parameters, we find that high redshift data is more consistent with Planck.}

Forthcoming releases from the Dark Energy Spectroscopic Instrument (DESI)~\cite{DESI:2016fyo} will be in a position to confirm or refute our assumption iv). Thus, our result is preliminary. Our findings also provide a targeted prediction for tomographic studies of cosmic shear \cite{HSC:2018mrq, KiDS:2020suj, DES:2021wwk}. These surveys, i. e. HSC, KiDS and DES, currently employ redshift bins with varying effective redshifts across the surveys and shifts in cosmological parameters up to $1 \sigma$ are evident. Given our results here, it is plausible that $S_8$ {redshift} evolution will be detected as cosmic shear data quality improves through a better understanding of systematics. {If redshift evolution is not detected, and assuming $S_8$ tension is physical, then there must be evolution with scale \cite{ACT:2023ipp}.} Finally, our results caution theoretically that physics that alters the radius of the sound horizon~\cite{Knox:2019rjx} cannot account for these \textit{expected} hallmarks of model breakdown.  

\section*{Acknowledgements}
We thank Dragan Huterer, Gabriela Marques, Leandros Perivolaropoulos, Sunny Vagnozzi, and Matteo Viel for correspondence on related topics. SAA is funded by UGC non-NET Fellowship scheme of Govt. of India. \"{O}A acknowledges the support by the Turkish Academy of Sciences in the scheme of the Outstanding Young Scientist Award (T\"{U}BA-GEB\.{I}P). \"{O}A is supported in part by TUBITAK grant 122F124. SP acknowledges hospitality of Bu-Ali Sina University while this work carried out. MMShJ and SP are supported in part by SarAmadan grant No ISEF/M/401332. AAS acknowledges the funding from SERB, Govt of India under the research grant no: CRG/2020/004347. This article/publication is based upon work from COST Action CA21136 – “Addressing observational tensions in cosmology with systematics and fundamental physics (CosmoVerse)”, supported by COST (European Cooperation in Science and Technology).

\section*{Data Availability}
All the data analysed in this study are in the public domain and references are provided. 

\bibliographystyle{mnras}
\bibliography{refs}

\bsp	
\label{lastpage}

\appendix

\section{Larger Archival Data Set}
\begin{figure}
\includegraphics[width=80mm]{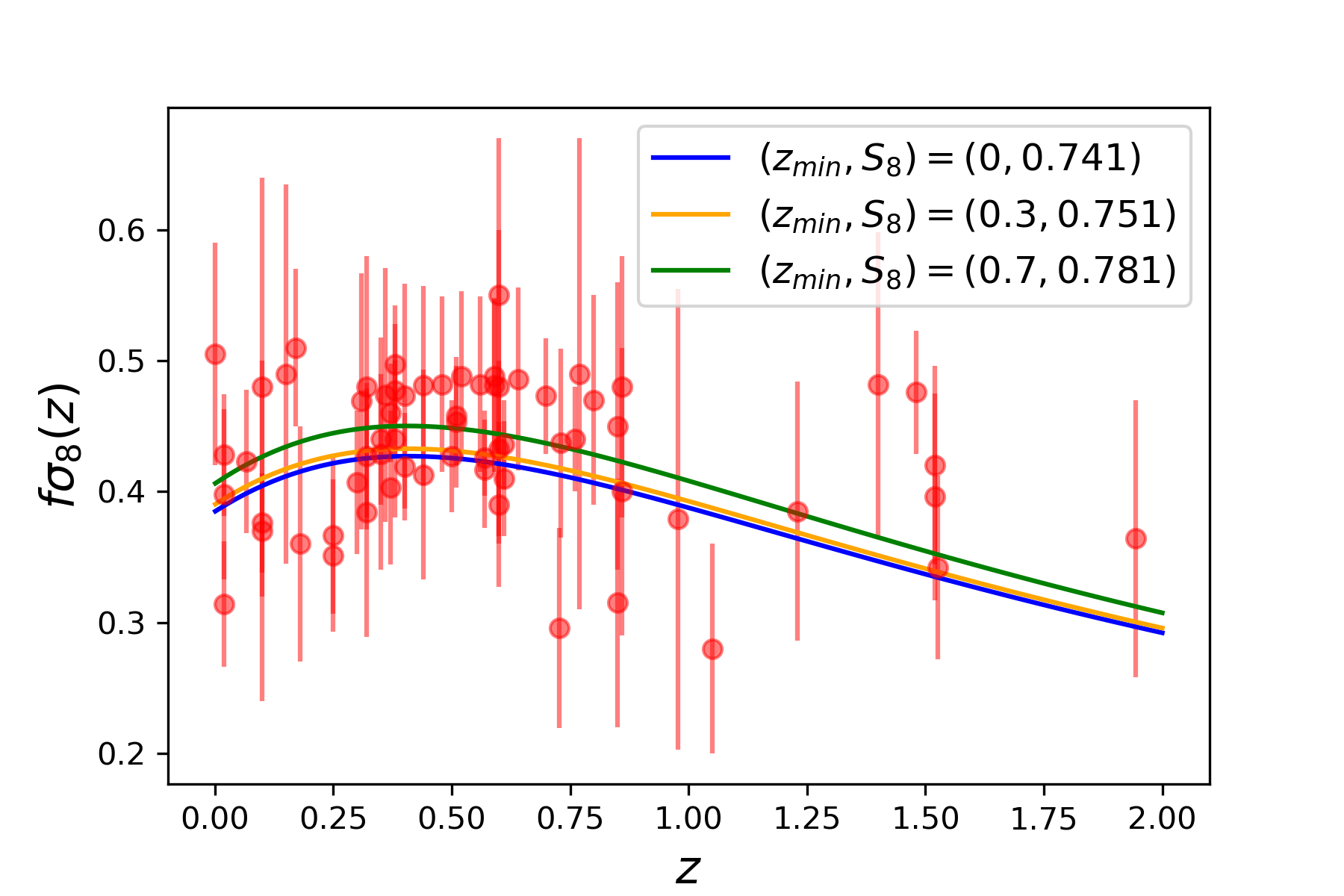} 
\caption{$f \sigma_8(z)$ constraints 
alongside best fit values of $S_8$ from fitting (\ref{fs8}) in the range $z \geq z_{\textrm{min}}$. Throughout we employ the Gaussian prior $\Omega_m = 0.3111 \pm 0.0056$.}
\label{fig:leandros_data}
\end{figure}

Given that the increase in $S_8$ {in the text} is driven exclusively by two high redshift data points, it is prudent to work with a larger data set. The role of the second data set is to provide a sanity check, since the larger data set comes with caveats that we discuss below. To that end, we focus on Table II of~\cite{Kazantzidis:2018rnb}, where one finds 63 historical measurements of $f \sigma_8(z)$~\cite{Beutler:2012px, Blake:2013nif, Howlett:2014opa,  Okumura:2015lvp, Huterer:2016uyq, Song:2008qt, Blake:2012pj, Davis:2010sw, Hudson:2012gt, Turnbull:2011ty, Samushia:2011cs, Tojeiro:2012rp, delaTorre:2013rpa, Chuang:2012qt, Sanchez:2013tga, Wang:2017wia, Feix:2015dla, Chuang:2013wga, BOSS:2016wmc, BOSS:2016psr, Wilson:2016ggz, Gil-Marin:2016wya, Hawken:2016qcy, delaTorre:2016rxm, Pezzotta:2016gbo, Feix:2016qhh, Howlett:2017asq, Mohammad:2017lzz, Shi:2017qpr, Gil-Marin:2018cgo, Hou:2018yny, Zhao:2018gvb}. We add more recent data from~\cite{Bautista:2020ahg, deMattia:2020fkb, Neveux:2020voa} to increase the sample to 66 data points. We observe that Refs.~\cite{Beutler:2012px, Blake:2013nif, Howlett:2014opa, Okumura:2015lvp,  Huterer:2016uyq} are common to both data sets in Fig.~\ref{fig:dragan_data} and Fig.~\ref{fig:leandros_data}. The other data points are the work of independent groups, but as cautioned in~\cite{Kazantzidis:2018rnb}, most analyses of growth rate data, e.g.~\cite{Nguyen:2023fip}, tend to pick 20 odd data points to counteract any overcounting of data. However, this choice of \textit{independent} data points is largely subjective. Moreover, different studies make use of different fiducial cosmologies. To address this latter concern, we follow equation (2.6) of \cite{Nesseris:2017vor} (see also~\cite{Macaulay:2013swa}) and  correct for this choice of fiducial model by performing a rescaling by the combination $H(z) D_{A}(z)$, where $H(z)$ and $D_{A}(z)$ denote the Hubble parameter and angular diameter distance at a given $z$. {This means that the data set is consistent with the $\Omega_m$ prior by construction. Note, following \cite{Nguyen:2023fip}, we did not perform this correction {in the body of the letter}. In the $f \sigma_8(z)$ literature, one finds groups who correct and do not correct for the fiducial cosmology.}  

\begin{figure}
\includegraphics[width=80mm]{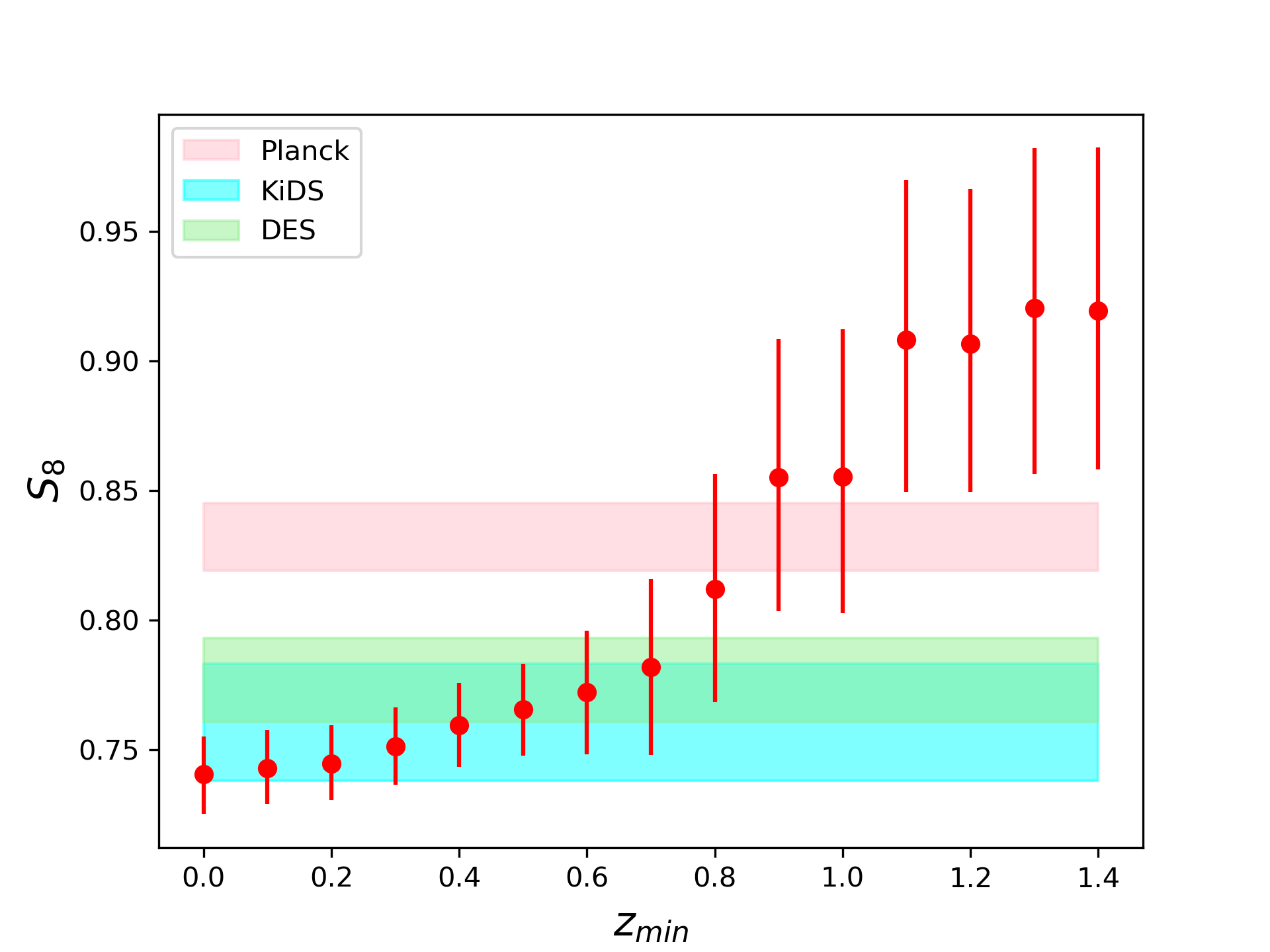} 
\caption{Removing $f \sigma_8(z)$ constraints with $z < z_{\textrm{min}}$ while fitting expressions for the $\Lambda$CDM model (\ref{fs8}) subject to a Gaussian prior $\Omega_m = 0.3111 \pm 0.0056$. An increasing $S_8$ trend with effective redshift is evident and the error bars correspond to $1 \sigma$ confidence intervals. Constraints from KiDS, DES and Planck are illustrated for comparison.}
\label{fig:leandros_data_trend}
\end{figure}

In Fig.~\ref{fig:leandros_data_trend} we show the effect of removing $f \sigma_{8}(z)$ constraints below $z = z_{\textrm{min}}$. Where relevant, {e.g. WiggleZ ~\cite{Blake:2012pj}}, we crop covariance matrices accordingly as we remove data points. Our strong Gaussian prior on $\Omega_{m}$ ensures that our \textit{a posteriori} distributions from MCMC analysis remain Gaussian. As a result, the $1 \sigma$ errors in Fig.~\ref{fig:leandros_data_trend} are representative. The reader will note that, in line with expectations, the size of the error bar increases as we remove data. Furthermore, the error bars in Fig.~\ref{fig:leandros_data_trend} are correlated, because all plotted constraints larger than a given $ z_{\textrm{min}}$ share data points.  

\begin{figure}
\includegraphics[width=80mm]{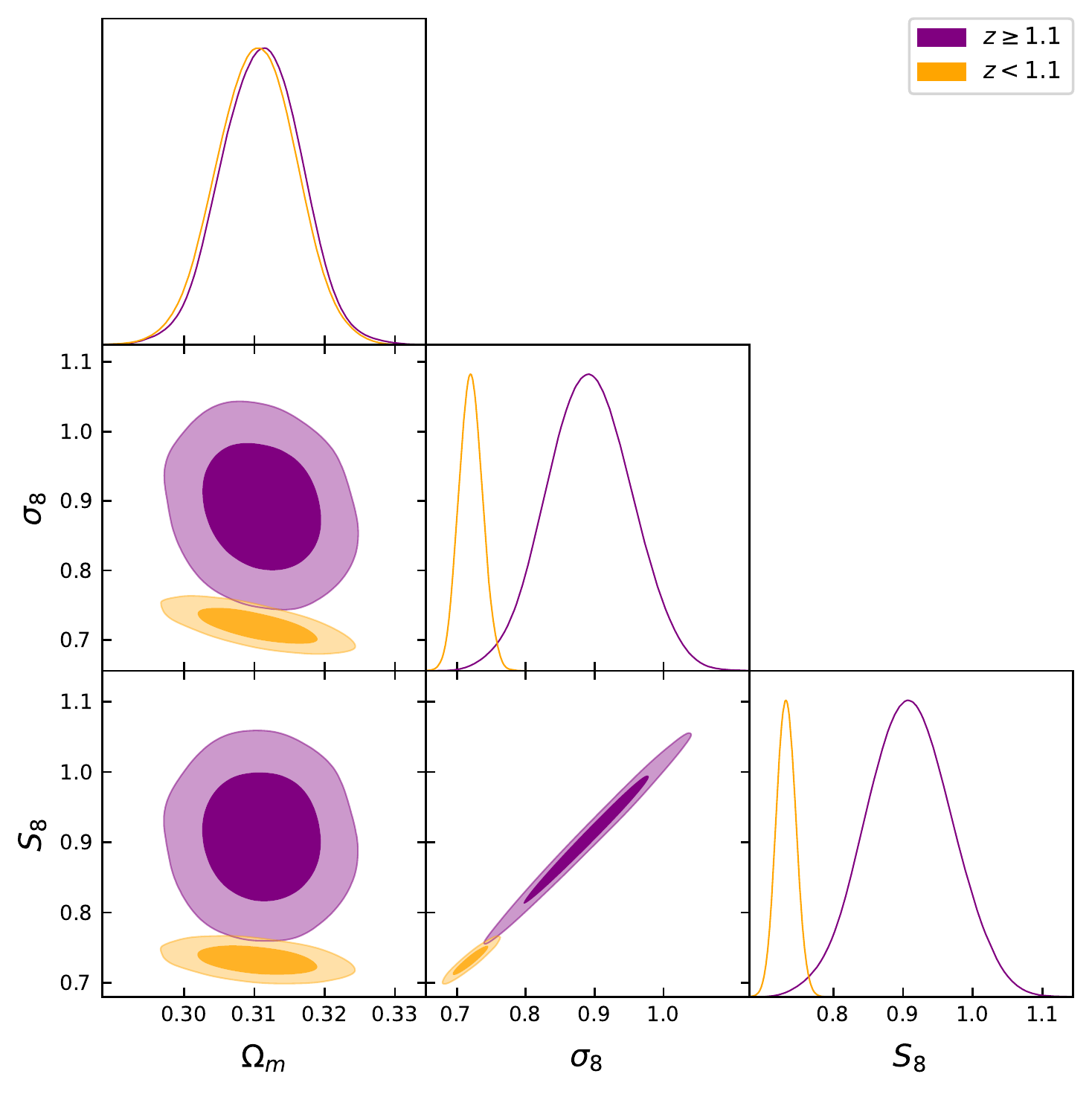} 
\caption{Marginalisation of parameters $(\Omega_m, \sigma_8, S_8)$ with Gaussian prior $\Omega_m = 0.3111 \pm 0.0056$ and compilation of $f \sigma_8(z)$ data
split at $z = 1.1$. $S_8$ is reconstructed from $(\Omega_m, \sigma_8)$ MCMC chains. The discrepancy in the $S_8$ plane between low and high redshift subsamples is {$2.8 \sigma$}.}
\label{fig:leandros_prior}
\end{figure}

We can address criticism regarding the correlated constraints in Fig.~\ref{fig:leandros_data_trend} by simply splitting the sample into independent subsamples. Our results in Fig.~\ref{fig:leandros_data_trend} suggest that any split at $z \sim 1$ may maximise the discrepancy between low redshift and high redshift subsamples. {Once again, the trend is directly visible in the data in Fig. \ref{fig:leandros_data}, since 7 from 8 data points above $z =1$ prefer larger values of $S_8/\sigma_8$ than the full sample ($z_{\textrm{min}} = 0$).} The choice of split is somewhat arbitrary, as it depends on the composition or makeup of the overall sample, notably redshift distribution and quality of the data. For this reason, the split here is conducted at $z = 1.1$, whereas it was $z = 0.7$ in the {body of the letter}. Note that our sample here now has an extended range, so we can explore higher redshift splits of the sample. Concretely, we find that a split at $ z= 1.1$ leads to {$S_8 = 0.732 \pm 0.014$ at lower redshifts and $S_8 = 0.908 \pm 0.062$ at higher redshifts, with a tension at $2.8 \sigma$ as illustrated in Fig. \ref{fig:leandros_prior}. Once again, it is worth noting that $S_8$ from data below $z=1.1$ is in tension with Planck at $\sim 5 \sigma$, while the data above $z =1.1$ is consistent with Planck at $\sim 1 \sigma$.} 

\begin{figure}
\includegraphics[width=80mm]{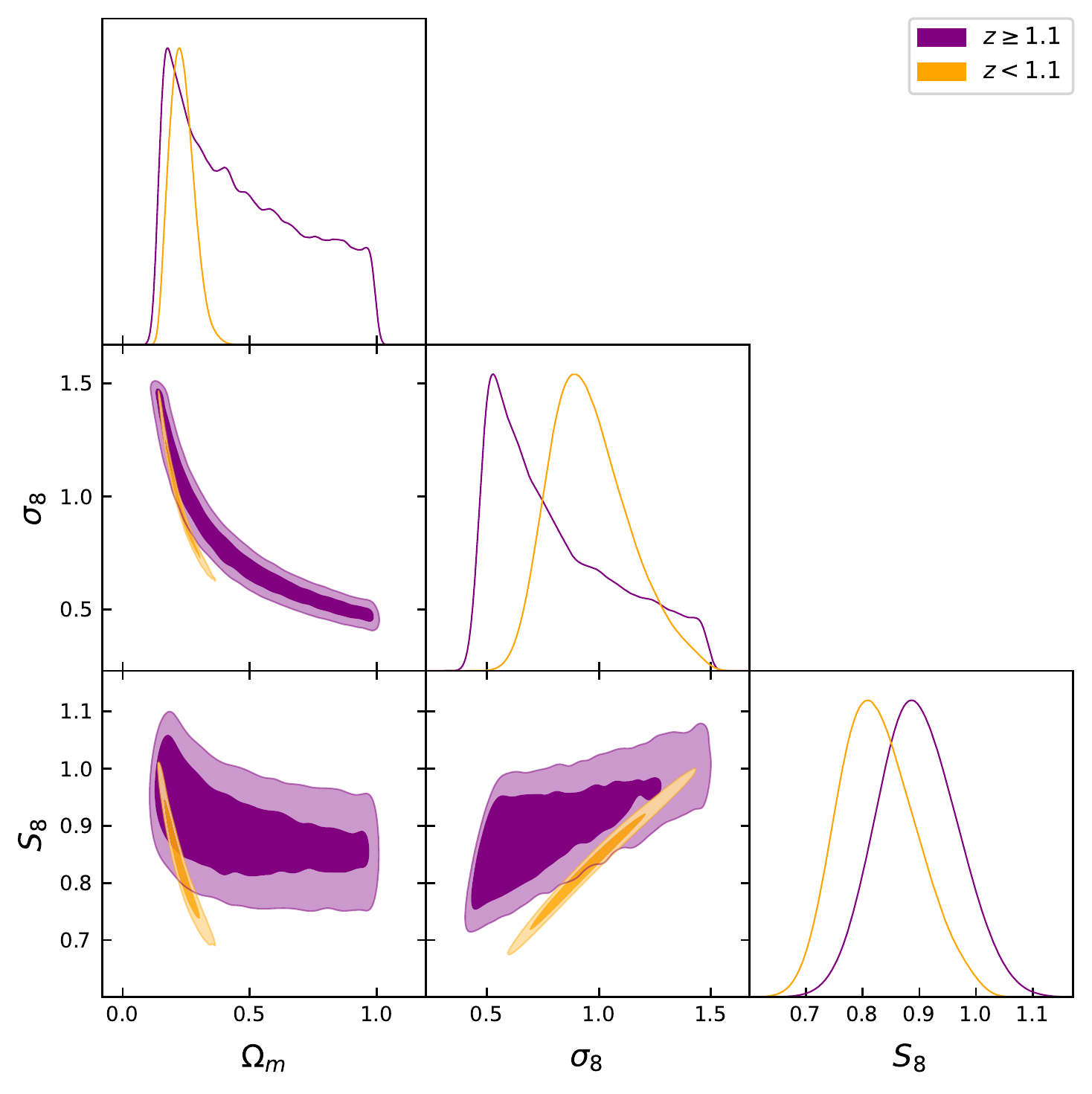} 
\caption{{Same as Fig. \ref{fig:leandros_prior} but without the CMB+BAO $\Omega_m$ prior. The 2D posterior in the $(\Omega_m, \sigma_8)$-plane points to a degeneracy that is poorly broken by higher redshift constraints since the $1 \sigma$ confidence intervals saturate the bounds.}}
\label{fig:leandros_no_prior}
\end{figure}

{Removing the $\Omega_m$ prior, the corresponding results are $(\Omega_m, S_8) = (0.229^{+0.051}_{-0.044}, 0.823^{+0.077}_{-0.064})$ $ (z < 1.1)$ and $(\Omega_m, S_8) = (0.435^{+0.352}_{-0.227}, 0.894^{+0.075}_{-0.068})$ $(z \geq 1.1)$. Evidently, all inferences are now consistent with Planck within $2 \sigma$, however as can be seen from Fig. \ref{fig:leandros_no_prior}, there is degeneracy in the $(\Omega_m, \sigma_8)$-plane that the data fails to break. Thus, care needs to be taken with these inferences as they will be prior dependent. Noting that $\Omega_m$ is lower than expected in the lower redshift subsample, it is interesting to studying the full sample without the $\Omega_m$ prior. Doing, so we find $(\Omega_m, \sigma_8, S_8) = (0.191^{+0.035}_{-0.029}, 1.118^{+0.185}_{-0.158}, 0.890^{+0.066}_{-0.059})$. Evidently there is some discrepancy with Planck, whereby $\Omega_m,  \sigma_8$ and $S_8$ are respectively $3.5 \sigma, 1.9 \sigma$ and $1 \sigma$ away. In both data sets considered in this work, there is evidently tension with Planck in one of the parameters.} 

{Bearing in mind that not all the data points are independent, and as a result the errors are underestimated, the main take away from our analysis here is that the $S_8$ trend with Planck+BAO $\Omega_m$ prior is robust to changes in the data set. A secondary take away is confirmation that one expects to run into trouble in higher redshift bins without an $\Omega_m$ prior in line with equations (2) and (3).} 

\end{document}